\documentclass[12pt,aps,prd,superscriptaddress,
nofootinbib,
 amsmath,amssymb,
aps]{revtex4-2}

\usepackage{graphicx}
\usepackage{xcolor}
\usepackage{bm}

\newcommand{\Tv}{T_{\text{vis}}}
\newcommand{\Th}{T_{\text{hid}}}
\newcommand{\trh}{t_{\text{RH}}}
\newcommand{\Mp}{M_{\text{pl}}}
\newcommand{\Mpe}{\tilde{M}_{\text{pl}}}

\newcommand{\comments}[1]{}

\usepackage{hyperref}
\usepackage{cleveref}

\begin{document}

\preprint{APS/123-QED}

\title{Hidden Sector Relic Abundances from an Effective Planck Mass}

\author{Donald Liveoak}
\affiliation{Leinweber Institute for Theoretical Physics, Physics Department, University of Michigan, Ann Arbor, MI 48109, USA
}

\author{Anshuman Maharana}
\affiliation{Leinweber Institute for Theoretical Physics, Physics Department, University of Michigan, Ann Arbor, MI 48109, USA
}
\affiliation{Harish-Chandra Research Institute, Chhatnag Road, Jhunsi,
Prayagraj, Uttar Pradesh 211019, India}

\affiliation{Homi Bhabha National Institute, Training School Complex, Anushakti Nagar, Mumbai
400094, India}

\author{James D. Wells}%
\affiliation{%
 Leinweber Institute for Theoretical Physics, Physics Department, University of Michigan, Ann Arbor, MI 48109, USA
}

\date{\today}

\begin{abstract}
{\it Abstract:}  Hidden sectors are generic in beyond the Standard Model constructions and string compactifications. We consider models featuring hidden sectors that have dark matter candidates. The relic density is set in part by the energy-injecting inflaton branching ratio $B$ to the sector. We develop a general framework to determine the dependence on $B$ for thermalized hidden sectors by defining an effective Planck mass for the Boltzmann equations for the hidden sector degrees of freedom. We apply this framework to several representative examples, including freeze-out/freeze-in WIMPs and multi-component dark matter. We find that the fractional energy density carried by freeze-out relics (in both single and multi-component models) and monopoles produced from second order phase transitions scale as $B^{1/4}$ and thus may lead to an overclosure problem, even with $B \ll 1$. On the other hand, the energy density of freeze-in relics scales as $B^{5/4}$ and is thus cosmologically irrelevant for sufficiently small $B$. Finally, we discuss the cumulative implications of a large number of hidden sectors.

\end{abstract}

\maketitle

\section{Introduction}\label{sec:intro}

Over the last two decades, observations have put the existence of dark matter (DM) on a firm footing. Despite this, a microscopic description of DM remains elusive. There are a plethora of theoretical candidates which can serve as DM in our Universe (see, e.g. \cite{bertone2005particle, feng2010dark, cirelli2025darkmatter} for reviews). This wide range of possibilities necessitates the development of generic tools that can help provide a systematic understanding of the space of models. In this article, we will take steps in this direction in the context of \textit{hidden sector DM models}, where DM resides in a hidden sector with a gauge group independent of the Standard Model.

Hidden sectors appear in various beyond the Standard Model theories. They also naturally arise in string compactifications, where they are often needed for mathematical consistency (see,  for example, \cite{Giedt:2000bi, Dienes:2007ms, Anderson:2013xka} for heterotic constructions, \cite{Cvetic:2004ui, Gmeiner:2005vz, Loges:2022mao, Marchesano:2024gul} for  intersecting D-brane models,  \cite{Taylor:2015ppa} for F-theory models,  \cite{Dijkstra:2004cc} for rational conformal field theory constructions, and \cite{Acharya:2016fge} for M-theory models). It is common for models to contain a large number of hidden sectors. In addition, even within a particular model, one can expect hidden sectors with distinct fundamental scales due to warping in the extra dimensions \cite{Hebecker:2006bn}.  These hidden sectors can possess particles which are stable on cosmological timescales and may serve the role of DM. The conserved quantum number can be associated with discrete and continuous isometries of the geometry, unbroken subgroups of gauge symmetries, or charges of non-perturbative objects (see e.g. \cite{Witten:1985xc, Shiu:2003ta, Lebedev:2006kn, Lebedev:2007hv, Frey:2009qb, Berasaluce-Gonzalez:2011gos,  Halverson:2016nfq, Halverson:2018vbo, Cicoli:2023opf}). 

The flexibility of hidden sector DM is attractive from a model-building perspective. On the other hand, these models can pose challenges to the well-established cosmological history. Cosmology thus serves as a powerful tool for constraining the space of phenomenologically viable models.

The fractional energy density carried by a hidden sector at early times is an important quantity for determining the role that the sector plays in the cosmological history. In the inflationary context, this is determined by the branching ratio $B$ of the inflaton into the hidden sector. The naive expectation is that sectors with low values of the fractional energy density are irrelevant for cosmology. This is true for dark radiation -- the contribution of a hidden sector to $\Delta N_{\rm eff}$ (the effective number of light neutrino-like species) is proportional to its fractional energy density\footnote{This constrains $B$ to be much less than one. The requirement is often phrased as asymmetric reheating, see e.g. \cite{Hodges:1993yb, Berezhiani:1995am, Adshead:2016xxj, Sandick:2021gew}.}. However, this is not true in general. Recently, it was found that the contribution of dark monopoles to the DM relic density scales as $B^{1/4}$ \cite{Liveoak:2026muy}. Thus, overclosure constraints from dark monopoles are practically insensitive to the inflaton branching ratio, unless it is anomalously low. A recent analysis provided constraints
on hidden sector symmetry breaking above the 100 PeV scale \cite{Liveoak:2026muy}. The case of dark monopoles clearly exhibits that early time fractional energy density is central to the phenomenology of models with hidden sectors. 

Here, we focus on hidden sectors which thermalize internally and are radiation-dominated in the early Universe. We develop a general formalism to compute the dependence of the hidden sector relic abundance on the fraction of energy it carries at early times. We find that this dependence is captured by considering an {\it{effective Planck mass}} for cosmological evolution of hidden sector abundances. We apply our formalism to representative freeze-in and freeze-out scenarios, both for single-component and multi-component DM models. Interestingly, we find that for the freeze-out models, large regions of parameter space have relic abundance $\propto B^{1/4}$, as in the case of dark monopoles. This implies that hidden sectors which undergo freeze-out can contribute significantly to the energy density of the Universe, leading to overclosure even in the case where $B \ll 1$. On the other hand, for freeze-in models, we find that the relic density $\propto B^{5/4}$; thus, overclosure may be avoided in this case for small $B$.

Our results have important implications for model building in string phenomenology. We find that model building requires detailed understanding of the inflaton branching ratio to all sectors (even if they are minute) so that relic densities can be reliably computed. As in the case of dark monopoles, overclosure considerations provide strong bounds on hidden sector scales.

This paper is structured as follows. In Section~\ref{sec:general}, we outline our general analytical argument for the computation of relic densities via an effective Planck mass. In Section~\ref{sec:applications}, we apply our analytical understanding to a variety of representative freeze-out and freeze-in models and confirm our results with numerical integrations. In Section~\ref{sec:discussion}, we generalize our analysis to the case of several hidden sectors and highlight the implications to hidden-sector model building. We conclude in Section~\ref{sec:conclusion}.

\section{General analysis}
\label{sec:general}
We consider the case of a single hidden sector which is thermally isolated from the visible sector. If the branching ratio of the inflaton into the hidden sector is $B$, then the ratio of the energy densities of the hidden and visible sectors is 
\begin{equation}
    \frac{\rho_{\text{hid}}(\trh)}{\rho_{\text{vis}}(\trh)} = \frac{B}{1-B},
\end{equation}
where $\trh$ is the time of reheating. Assuming that both sectors are radiation-dominated and the entropy densities $s_{\rm hid}$ and $s_{\rm vis}$ are independently conserved, we may solve for the ratio of the hidden and visible sector temperatures
\begin{equation}
    \frac{\Th(t)}{\Tv(t)} = \left( \frac{g_\ast^{\rm hid}(\trh)}{
 g_\ast^{\rm vis}(\trh)} \right)^{1/12}
 \left( \frac{g_\ast^{\rm vis}(t)}{
 g_\ast^{\rm hid}(t)} \right)^{1/3}  \left( \frac{B}{1-B} \right)^{1/4},
\end{equation}
where $g_\ast^{\rm hid}(t)$ and $g_\ast^{\rm vis}(t)$ are the effective number of relativistic hidden sector and visible sector degrees of freedom, respectively (1 per bosonic d.o.f., $7/8$ per fermionic d.o.f.).

If $B\sim \mathcal{O}(0.1)$ or higher, the hidden sector energy density is commensurate to that of the visible sector. In this case, the hidden sector is severely constrained by Planck observations of the number of relativistic degrees of freedom  $\Delta N_{\rm eff}$ \cite{Planck:2018vyg}. Thus, we specialize to the limit $B \ll 1$. In this case, the energy density of the Universe is dominated by that of the visible sector, so the Hubble parameter depends only on the visible sector temperature via $H=\Tv^2/C\Mp$ where $C=\sqrt{90/\pi^2 g_\ast^{\rm vis}}$ and $\Mp \approx 2.4 \times 10^{18}\, \text{GeV}$ is the reduced Planck mass. Additionally, we assume that $g_\ast^{\rm hid}(t)$ and $g_\ast^{\rm vis}(t)$ remain approximately constant in the relevant era of cosmological evolution. Furthermore, to reduce clutter in our expressions, we suppress coefficients $\sim g_\ast^{\rm hid}/g_\ast^{\rm vis}$ which may easily be restored by counting powers of $B$. Thus, in this framework, the ratio between the hidden and visible sector temperatures is $\Th / \Tv = B^{1/4}$.

Suppose the hidden sector has $K$ constituents, each with number density $n_i$. The evolution of each species is determined by the set of coupled Boltzmann equations
\begin{equation}
    \dot{n}_i+3Hn_i = \mathcal{C}_i[\{n_j\}, \Th].
\end{equation}
The term $\mathcal{C}_i$ encodes the microscopic processes which may produce or remove particles of species $i$ from the Universe, such as pair annihilation, decay, and their reverse processes. In general, $\mathcal{C}_i$ may be a complicated function which depends on the number density of each particle which interacts with species $i$, as well as the hidden sector temperature $\Th$. Importantly, since the hidden sector is decoupled from the visible sector, $\mathcal{C}_i$ is independent of visible sector degrees of freedom. Thus, the only dependence of the evolution of $n_i$ on the visible sector is through the expansion term $3Hn_i$.

It is convenient to work in terms of hidden-sector comoving abundances $Y_i^{\rm hid} \equiv n_i/s_{\rm hid}$. The Boltzmann equations may then be written
\begin{equation}
    \frac{dY_i^{\rm hid}}{dx_{\rm hid}} = \frac{\mathcal{C}_i[\{n_j\}, \Th]}{Hx_{\rm hid} s_{\rm hid}} = \frac{C\Mp B^{1/2}}{\Th^2 s_{\rm hid}x_{\rm hid}} \mathcal{C}_i[\{n_j\}, \Th] \equiv \Mp B^{1/2} \widehat{\mathcal{C}}_i[ \{Y^{\rm hid}_j \}, x_{\rm hid}],
\end{equation}
Here, we define $x_{\rm hid}=m/\Th$ where $m$ is a hidden sector mass scale. We define an {\it effective Planck mass} 
\begin{equation}
\label{effectivePlanck}
\Mpe \equiv \Mp B^{1/2},
\end{equation}
with which  the Boltzmann equations become
\begin{equation}
\label{eq:boltfinal}
    \frac{dY_i^{\rm hid}}{dx_{\rm hid}} = \Mpe \widehat{\mathcal{C}}_i[ \{Y^{\rm hid}_j \}, x_{\rm hid}]
\end{equation}
Note that this  has exactly the same structural form as the usual \textit{visible sector} Boltzmann equations. Therefore, the distinction between the hidden and visible sector temperatures is accounted for simply via a redefinition of the Planck mass. A few comments are in order:

\begin{itemize}

\item The branching ratio appears in the differential equations \eqref{eq:boltfinal} only via the effective Planck mass $(\Mpe)$.
Thus, the dependence of asymptotic abundances on $B$ is captured in either via $\Mpe$, or via the $B$-dependence of the initial conditions.

\item In a large class of models (such as freeze-out), the initial conditions are set by thermality; as such, the asymptotic abundances depend weakly on the initial conditions. In such cases, the dependence on the branching ratio is determined by the effective Planck mass dependence of the solutions. Furthermore, expanding the solution in powers of $\tilde{M}_{\rm pl}$ can be used to determine the leading order dependence in $B$.

\item There can be situations (e.g., monopole production during a second order phase transition \cite{Liveoak:2026muy}) where the initial conditions have dependence on $B$. In such cases, both the initial conditions and $\Mpe$ can be relevant for determining the branching ratio dependence of the final abundance.

\item The visible sector asymptotic abundances $(Y_{i, \infty}^{\rm vis})$ are related to the hidden sector abundances by an entropy
ratio conversion factor
$$
  Y_{i, \infty}^{\rm vis} =   { s_{\rm hid} (t_*)\over s_{\rm vis} (t_*) } Y_{i, \infty}^{\rm hid},
$$

with the ratio of entropies evaluated at any convenient time. This generally introduces powers of $B$ and $g_\ast^{\rm hid}/g_\ast^{\rm vis}$. For a power law scaling of the
asymptotic hidden sector abundance,  $Y_{i,\infty}^{\rm hid} \propto (\tilde{M}_{\rm pl})^n$,
$$
  \Omega_i \sim {1 \over (g_*^{\rm vis})^{1/4} (g_*^{\rm hid})^{3/4 +n/2} } B^{3/4 + n/2}
$$
Finally, one can use $Y^{\rm vis}_{i, \infty}$ to compute the fractional relic densities

\begin{equation}
    \Omega_i = \frac{\rho_i}{3H_0^2 \Mp^2} = \frac{m_i s_0^{\text{vis}} Y_{i,\infty}^{\rm vis}}{3H_0^2 \Mp^2},
\end{equation}
where $\rho_i$ is the present-day energy density of the $i^{\rm th}$ relic, $m_i$ its mass, $H_0$ is the Hubble constant, and $s_0^{\rm vis}$ is the present-day entropy density of  visible sector. 

\end{itemize}

\section{Applications}\label{sec:applications}
We now apply our analytical framework to a variety of DM models. We start by considering single-component DM and then move on to multi-component models. For each single-component model, we briefly review the analytical calculation for the asymptotic abundance in the visible sector. We apply our formalism to calculate the asymptotic abundance for the corresponding hidden-sector model, and confirm the $B$-dependence by numerically integrating the Boltzmann equation.

\subsection{Single-component freeze-out}\label{sec:freezeout}
Weakly-interacting massive particles (WIMPs) are a broad and well-motivated class of stable DM particles which have mass $m_\chi \sim 10\, \text{GeV} - 10\, \text{TeV}$ \cite{feng2010dark}. WIMPs may annihilate into SM particles via the process $\chi \chi \to \rm SM$. Here, we briefly review the so-called \textit{WIMP miracle}, in which the correct present-day abundance of DM is obtained via a freeze-out process involving a weak-scale annihilation cross section.

The Boltzmann equation for the comoving abundance of visible sector WIMPs, $Y_\chi$, is
\begin{equation}
    \frac{dY_\chi}{dx} = -\frac{s\langle \sigma v\rangle_{\chi\chi\to \rm SM}}{Hx}\left(Y_\chi^2-Y_{\chi, \rm eq}^2(x)\right),
\end{equation}
where $x=m_\chi/\Tv$, $\langle\sigma v\rangle_{\chi\chi\to \rm SM}$ is the thermally-averaged cross section of annihilations of the WIMPs to the SM, and the equilibrium abundance
\begin{equation}
    Y_{\chi, \rm eq}(x) = \frac{45}{4\pi^4} \frac{g_\chi}{g_\ast^{\rm vis}} x^2 K_2(x),
\end{equation}
where $g_\chi$ is the number of WIMP spin states, and $K_2(x)$ is the modified Bessel function of the second kind. We assume an $s$-wave electroweak annihilation so that
\begin{equation}
    \langle\sigma v\rangle_{\chi\chi\to \rm SM} = \frac{\alpha_{\rm weak}^2}{m_\chi^2},
\end{equation}
where $\alpha_{\rm weak}^{-1}\sim 30$ is the fine structure constant of the weak interaction. It is convenient to collect the cosmological and annihilation parameters and define
\begin{equation}
    \lambda \equiv \frac{sx}{H} \langle\sigma v\rangle_{\chi\chi\to \rm SM} = \frac{2\pi^2}{45}g_\ast^{\rm vis} \Tv^3 \frac{m_\chi}{\Tv}  \sqrt{\frac{90}{\pi^2 g_\ast^{\rm vis}}} \frac{\Mp}{\Tv^2} \frac{\alpha_{\rm weak}^2}{m_\chi^2} = \sqrt{\frac{8\pi^2}{45} g_\ast^{\rm vis}} \alpha_{\rm weak}^2 \frac{\Mp}{m_\chi},
\end{equation}
so that the Boltzmann equation simplifies to
\begin{equation}\label{eq:boltzmann-wimp}
    \frac{dY_{\chi}}{dx} = -\frac{\lambda}{x^2} \left(Y_\chi^2-Y_{\chi, \rm eq}^2(x)\right).
\end{equation}
At early times (and high temperatures), $x\ll 1$, so the right-hand side of \Cref{eq:boltzmann-wimp} forces the WIMPs to maintain their thermal abundance, $Y_\chi = Y_{\chi,\rm eq}(x)$. As the Universe cools, the interaction rate $\Gamma_{\chi \chi \to \rm SM}= Y_{\chi, \rm eq} (x) s_{\rm vis} \langle\sigma v\rangle_{\chi\chi\to \rm SM}$ will eventually be less than $H$; we call this point $x=x_{\rm fo}$. Then, the abundance of WIMPs is frozen out at the equilibrium abundance $Y_{\chi, \infty} = Y_{\chi, \rm eq}(x_{\rm fo})$.

The condition for freeze-out yields a transcendental equation in $x$ which gives $x_{\rm fo}\sim 20$ in typical scenarios \cite{feng2010dark}. The asymptotic abundance is 
\begin{equation}\label{eq:wimp-abundance}
    Y_{\chi,\infty} = \frac{H(x_{\rm fo})}{s(x_{\rm fo}) \langle\sigma v\rangle_{\chi\chi\to \rm SM}} = \frac{x_{\rm fo}}{\lambda} \propto \frac{m_\chi}{\Mp}.
\end{equation}

Now, we turn to the hidden sector analog of WIMP freeze-out, where $\chi'$ particles are neutral under the SM gauge group. The cosmological dynamics of freeze-out hidden sector WIMPs were first studied by Feng et al. \cite{Feng:2008mu} and subsequently in a variety of other scenarios \cite{Cheung:2010gj, Bringmann:2020mgx, Das:2010ts, Ackerman:2008kmp, Heikinheimo:2016yds, Berlin:2016vnh, Berlin:2016gtr, Nath:2024mgr, Reece:2015lch, Clery:2026fhj}. A freeze-out process similar to the visible sector case may occur, where $\chi'$ may annihilate to lighter hidden sector modes. Based on the scaling of $\Mp$ in \eqref{eq:wimp-abundance}, we expect the asymptotic abundance (in terms of the hidden sector entropy density) of $\chi'$ particles populated by freeze-out to be
\begin{equation}
    Y_{\chi',\infty}^{\rm hid} = \frac{x_{\rm fo}}{\lambda'} B^{-1/2},
\end{equation}
where $\lambda'$ is defined as before (still using $\Mp$), except with the annihilation cross section of the $\chi'$. Based on this, the fractional energy density is 
\begin{equation}
    \Omega_{\chi'} = \frac{m_{\chi'} s_0^{\rm vis} x_{\rm fo}}{3H_0^2 \Mp^2 \lambda'} B^{1/4},
\end{equation}
where the overall factor of $B^{1/4}$ includes the conversion from hidden sector to visible sector entropy densities. 

\begin{figure}
    \centering
    \includegraphics[width=1\linewidth]{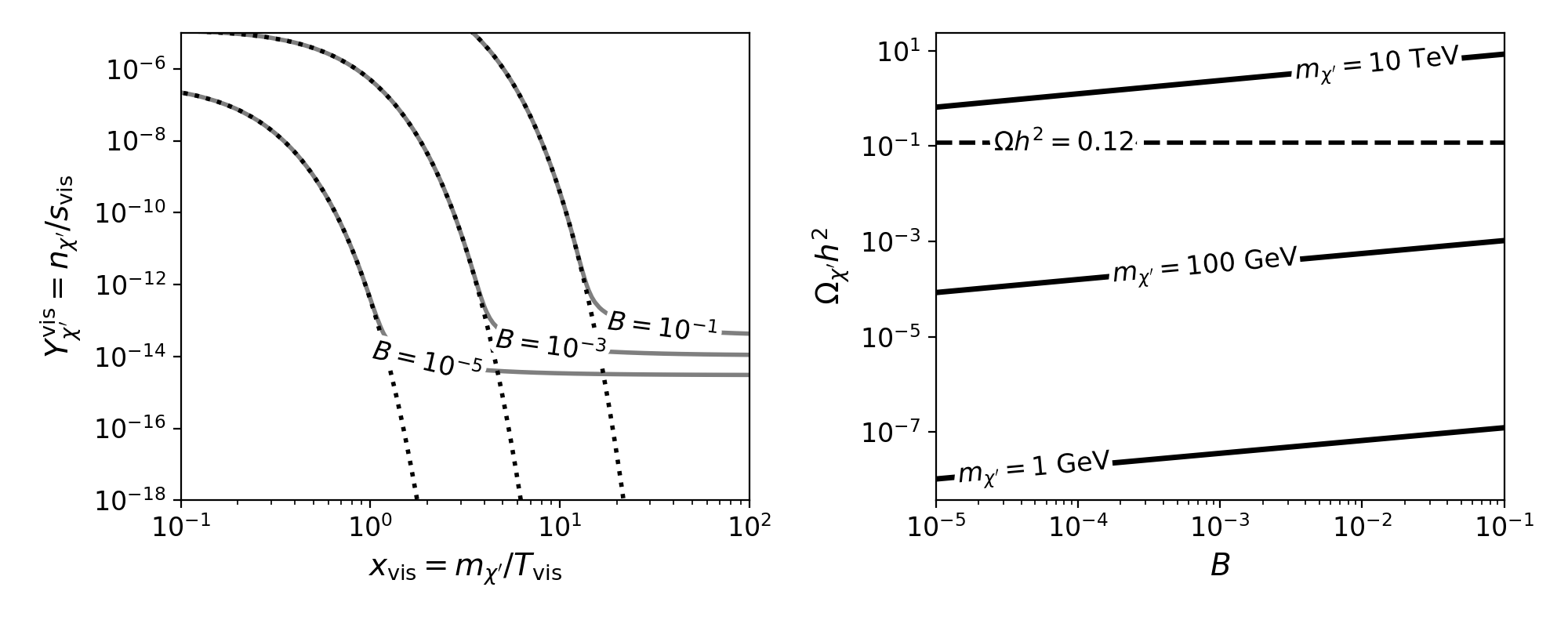}
    \caption{Left panel: Results of Boltzmann integration of hidden-sector WIMPs with $m_{\chi'}=100\, \rm GeV$, for $B=10^{-5}$, $B=10^{-3}$, and $B=10^{-1}$. The dashed lines indicate the corresponding equilibrium abundances. Right panel: present-day energy density of hidden sector freeze-out WIMPs as a function of $B$, for $m_{\chi'} = 1\, \rm GeV$, $m_{\chi'} = 100\, \rm GeV$, and $m_{\chi'}=10\, \rm TeV$. We numerically verify $\Omega_{\chi'}\propto B^{0.27}, B^{0.27}$ and $B^{0.28}$, respectively. The dashed line indicates the observed DM energy density, $\Omega h^2 \simeq 0.12$ \cite{Planck:2018vyg}.}
    \label{fig:wimp-freezeout}
\end{figure}

We numerically verify the $\sim B^{1/4}$ scaling of the energy density by numerically integrating the Boltzmann equation
\begin{equation}\label{eq:boltz-wimp-hidden}
    \frac{dY_{\chi'}^{\rm vis}}{dx_{\rm vis}} = -\frac{s_{\rm vis} \langle\sigma v\rangle_{\chi'\chi'\to \rm HS}}{Hx_{\rm vis}}\left((Y_{\chi'}^{\rm vis})^2 - Y_{\chi', \rm eq}^{\rm vis}(x_{\rm hid})^2\right),
\end{equation}
where $Y_{\chi'}^{\rm vis} \equiv n_{\chi'}/s_{\rm vis}$. For comparison with the visible sector case, we fix $\langle\sigma v\rangle_{\chi'\chi'\to \rm HS} = \alpha_{\rm weak}^2/m_{\chi'}^2$. We integrate from the strongly relativistic regime ($x_{\rm hid} = m_{\chi'}/\Th = 0.01$) until the abundance reaches its asymptotic value, typically at $x_{\rm hid} \sim \mathcal{O}(10)$.  The results of our integrations are shown in \Cref{fig:wimp-freezeout}, which confirm the freeze-out dynamics and $\sim B^{1/4}$ scaling of the energy density for $m_{\chi'} \in [1\, \rm GeV, 10\, \rm TeV]$ and $B \in [10^{-5}, 10^{-1}]$, up to sub-leading corrections due to the additional logarithmic dependence of $x_{\rm fo}$ on $B$ via $\Mp$ \cite{Kolb:1990vq}.

\subsection{Single-component freeze-in}

We now discuss the case of a single-component DM model which begins at a subthermal abundance but is rapidly populated via thermal production. This \textit{freeze-in} scenario has been studied in a variety of contexts, including Feebly Interacting Matter Particles (FIMPs) \cite{hall2010freeze, bernal2017dawn}, WIMPs diluted via entropy injection \cite{wang2023freeze}, dark photons \cite{dvorkin2019making}, and magnetic monopole production in high-temperature reheating scenarios \cite{turner1982thermal, Liveoak:2026dtx}.

We first review freeze-in dynamics in a visible sector model. We adopt the same simple WIMP model as in \Cref{sec:freezeout}, so that the visible sector Boltzmann equation is given by \Cref{eq:boltzmann-wimp}. However, we now assume $Y_{\chi}=0$ initially, so that the $\chi$ particles are populated thermally from the bath of SM particles, and choose $\lambda=10^{-6}$ to avoid rapid thermalization. In this case, $Y_\chi \ll Y_{\chi,\rm eq}$ for the entirety of the time when thermal production is efficient.\footnote{If this is not the case, i.e. for a stronger annihilation cross section, $Y_\chi$ will approach its equilibrium value, and we are left with the freeze-out scenario discussed before.}

The asymptotic abundance is then
\begin{equation}
    Y_{\chi, \infty} = \int_{x_i}^\infty dx\, \frac{\lambda}{x^2} Y_{\rm eq}^2(x) = \sqrt{\frac{8\pi^2}{45} g_\ast^{\rm vis}} \alpha_{\rm weak}^2 \frac{\Mp}{m_\chi} F(x_i),
\end{equation}
where $x_i=m_\chi/\Tv^i$ is the initial temperature parameter of the visible sector and $F(x_i) = \int_{x_i}^\infty dx\, Y_{\chi, \rm eq}^2(x)/x^2$ does not depend on the Planck mass. If the freeze-in occurs in the deep UV regime (i.e., $x_i \ll 1$), we may have that $\lambda F(x_i) \sim Y_{\chi, \rm eq}(x_i)$ in which case thermal production is strong enough for $\chi$ to reach its equilibrium abundance; in this case, the $\chi$ undergoes freeze-out. In the subsequent analysis, we neglect this possibility by specializing to the IR regime and choosing $x_i$ large enough so that the $\chi$ abundance remains comfortably subthermal.

We see that $Y_{\chi, \infty} \propto \Mp$, and thus we expect that the asymptotic abundance for the corresponding hidden sector model is obtained via the substitution $\Mp \to \Mp B^{1/2}$, yielding
\begin{equation}\label{eq:freezein-hidden-abundance}
     Y_{\chi', \infty}^{\rm hid} = \lambda' F(m_{\chi'}/\Th^i) B^{1/2},
\end{equation}
where $\Th^i$ is the initial temperature of the hidden sector. In realistic models, $\Th^i$ may also carry dependence on $B$, which is readily accounted for in \Cref{eq:freezein-hidden-abundance}. For simplicity, we assume that $\Th^i$ is independent of $B$. The fractional energy density is then
\begin{equation}
    \Omega_{\chi'} = \frac{m_{\chi'} s_0^{\rm vis}}{3H_0^2 \Mp^2} \lambda' F(m_{\chi'}/\Th^i)B^{5/4}.
\end{equation}

\begin{figure}
    \centering
    \includegraphics[width=\linewidth]{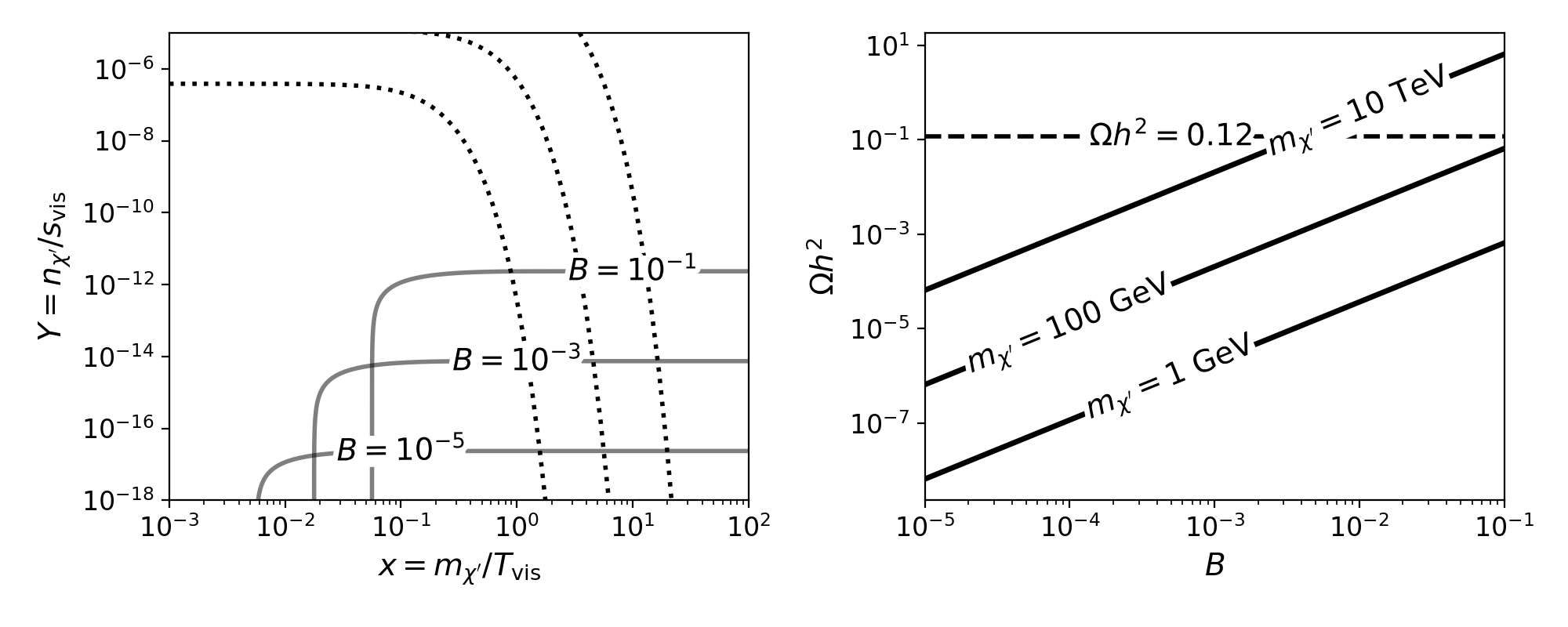}
    \caption{Same as \Cref{fig:wimp-freezeout}, except for a freeze-in scenario. We numerically verify $\Omega_{\chi'}\propto B^{1.25}$ in each case.}
    \label{fig:wimp-freezin}
\end{figure}

As before, we now confirm our results via numerical integration of the Boltzmann equation. The Boltzmann equation is again given by \Cref{eq:boltz-wimp-hidden}, but now we choose to start our integration at $x^i_{\rm hid}=0.1$ and $Y_{\chi'}^{\rm vis}(x^i_{\rm hid})=0$. We integrate until thermal production ceases (typically at $x_{\rm vis}\sim \mathcal{O}(1)$). We choose $m_{\chi'}\in[1\, \text{GeV}, 10\, \text{TeV}]$ and $B\in[10^{-5}, 10^{-1}]$. We numerically confirm that $\Omega_{\chi'}\propto B^{5/4}$ in each case. This confirms the results of our analytical framework in the case of freeze-in WIMPs.

Freeze-in production of very weakly interacting particles (such as axions, saxions) produced via non-renormalizable interactions \cite{Graf:2010tv, Graf:2012hb, garcia2020reheating} provides an exception to the above.  Such particles  can have early time dominated production  without being  thermalised. The number density evolution equation has the form:
\begin{equation}
\label{eq:nr}
  \dot{n} + 3H n  = {A_1 \over \Lambda^2} T^6_{\rm hid},
\end{equation}
where $\Lambda$ is the UV cut off for the non-renormalizable interaction and $A_1$ is set by hidden sector
couplings and parameters (we refer the reader to \cite{Graf:2010tv, Graf:2012hb} for its precise form). Integrating 
\eqref{eq:nr}, one finds:
$$
 {n \over T^{3}_{\rm hid}} \Bigg{|}_{\rm final} -   {n \over T^{3}_{\rm hid}}\Bigg{|}_{\rm initial}
 = {A_1 \Mp \over \Lambda^2} \sqrt{90/\pi^2 g_\ast^{\rm hid}} B^{1/2} \left( T_{\rm hid}^{\rm max} - T_{\rm hid}^{\rm stop} \right),
$$
where $\Th^{\rm stop}$ is the hidden sector temperature where the production stops.
The right hand side is dominated by the UV, therefore for a vanishing initial abundance we have:
$$
  Y_{\infty}^{\rm hid} \sim {A_1 \Mp \over \Lambda^2}  B^{1/2}  T_{\rm hid}^{\rm max}.
$$
For a fixed visible sector reheating temperature, $T_{\rm hid}^{\rm max} \sim B^{1/4} T^{\rm max}_{\rm vis}$. Thus $Y_{\infty}^{\rm hid} \sim B^{3/4}$, correspondingly $Y_{\infty}^{\rm vis} \sim B^{3/2}$.

\subsection{Multi-component DM}
We now turn to the case of models which have several interacting fields which may play the role of DM. For notational clarity, in this section, we will exclusively work in terms of abundances defined with respect to the visible sector entropy density and thus suppress the superscripts. 

\subsubsection{Unstable relics}
We consider scenarios in which several hidden sector relics are produced thermally. First, we consider the case of a heavy scalar relic $\Phi$ and light fermionic relic $f$ such that the decay process $\Phi \to f\bar f$ is kinematically allowed. We denote the mass of $\Phi$ and $f$ as $m_\Phi$ and $m_f$, respectively. 

Suppose the decay width of $\Phi \to f \bar{f}$ is $\Gamma$, and the thermally-averaged annihilation cross sections of $\Phi \bar\Phi$ and $f \bar f$ are $\langle\sigma_{\Phi\bar\Phi}v\rangle$ and $\langle\sigma_{f\bar f}v\rangle$. To describe the coupled evolution of $\Phi$ and $f$, the Boltzmann equation in terms of the respective (visible sector) comoving abundances $Y_\Phi$ and $Y_f$ is

\begin{align}
    \frac{dY_\Phi}{dx} &= -\frac{\langle\sigma_{\Phi\bar\Phi}v\rangle s_{\rm vis}}{Hx}\left(Y_\Phi^2 - Y_{\Phi,\text{eq}}^2(xB^{-1/4})\right) - \frac{\Gamma}{Hx} \left(Y_\Phi-\frac{Y_{\Phi,\text{eq}}(xB^{-1/4})}{Y^2_{f,\text{eq}}(xB^{-1/4})}Y_{f}^2\right)\\
    \frac{dY_f}{dx} &= -\frac{\langle\sigma_{f\bar f}v\rangle s_{\rm vis}}{Hx}\left(Y_f^2 -Y_{f,\text{eq}}^2(xB^{-1/4})\right) + \frac{\Gamma}{Hx} \left(Y_\Phi-\frac{Y_{\Phi,\text{eq}}(xB^{-1/4})}{Y^2_{f,\text{eq}}(xB^{-1/4})}Y_{f}^2\right),
\end{align}
where $x=m_\Phi/T_{\text{vis}}$. Here, we assume that the abundance of each particle is equal to that of its antiparticle. The equilibrium abundances $Y_{\Phi, \rm eq}$ and $Y_{f, \rm eq}$ depend on the hidden sector temperature but are defined with respect to the \textit{visible sector} entropy density. As before, the cosmological evolution only depends on the visible sector. We choose to parameterize these coupled ODEs as
\begin{align}
    \frac{dY_\Phi}{dx} &= -\frac{\lambda_{\Phi}}{x^2}\left(Y_\Phi^2 -Y_{\Phi,\text{eq}}^2(xB^{-1/4})\right) - \lambda_\Gamma x \left(Y_\Phi-\frac{Y_{\Phi,\text{eq}}(xB^{-1/4})}{Y^2_{f,\text{eq}}(xB^{-1/4})}Y_{f}^2\right)\\
    \frac{dY_f}{dx} &= -\frac{\lambda_f}{x^2}\left(Y_f^2 -Y_{f,\text{eq}}^2(xB^{-1/4})\right) + \lambda_\Gamma x \left(Y_\Phi-\frac{Y_{\Phi,\text{eq}}(xB^{-1/4})}{Y^2_{f,\text{eq}}(xB^{-1/4})}Y_{f}^2\right),
\end{align}
where
\begin{align}
    \lambda_{\Phi} &\equiv\frac{\langle\sigma_{\Phi\bar\Phi}v\rangle s_{\rm vis}}{H(\Tv=m_\Phi)} & \lambda_{f} &\equiv\frac{\langle\sigma_{f\bar f}v\rangle s_{\rm vis}}{H(\Tv=m_\Phi)}.
\end{align}
The parameters $\lambda_{\Phi}$ and $\lambda_{f}$ are dimensionless and characterize the strength of the (s-wave) annihilation processes, and $\lambda_\Gamma  \equiv \Gamma /H(\Tv=m_\Phi)$.

To demonstrate the applicability of our analytical results, we analyze three scenarios in this more complicated model. For each scenario, we first consider the corresponding visible sector model to gain a qualitative understanding of the dynamics. We then extend to the hidden sector case, demonstrating that the power-law dependence of the final abundances is consistent with our previous results. For simplicity, we fix $m_\Phi = 100\, \rm GeV$ and $m_f = 25\, \rm GeV$.

\textit{Decoupled freeze-out.} We first consider the case in which $\lambda_\Gamma = 0$ and the annihilation cross sections are consistent with WIMP freeze-out, $\lambda_\Phi m_{\Phi} =\lambda_f m_f = 3.7\times 10^{16} \, \rm GeV$. Additionally, we assume both relics begin at their equilibrium abundance. In this case, we simply have two decoupled relics undergoing freeze-out. 

For visible sector relics, we integrate from $x_{\rm vis} = 2$ until the relics reach their asymptotic abundance, which typically occurs at $x_{\rm vis} \sim \mathcal{O}(100)$. 

\begin{figure}
    \centering
    \includegraphics[width=\linewidth]{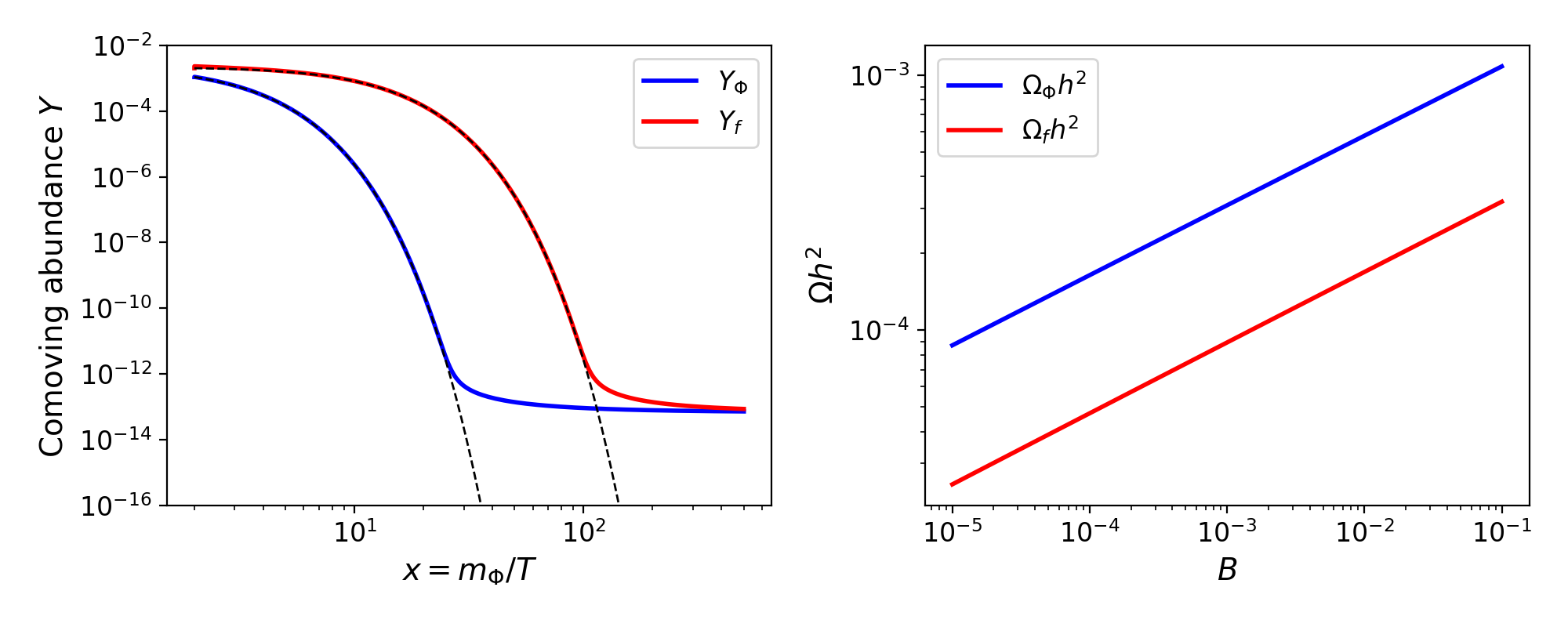}
    \caption{Left panel: Results of Boltzmann integration of decoupled visible sector WIMPs, $\Phi$ (blue) and $f$ (red). Freeze-out occurs at $x \sim 20$ for $\Phi$ and $x \sim 80$ for $f$. Right panel: Dependence of the final energy density of $\Phi$ (blue) and $f$ (red) on the inflaton branching ratio $B$. Both energy densities are proportional to $B^{0.28}$, in agreement with the case of single-component freeze-out.}
    \label{fig:decoupled-freezeout}
\end{figure}

The result of this integration is shown in the left panel of \Cref{fig:decoupled-freezeout}; as expected, both relics track their equilibrium abundance until annihilation becomes inefficient and the relics cannot maintain thermal equilibrium. This occurs at $x\sim 20$ for $\Phi$ and $x\sim 80$ for $f$, since $m_\Phi /m_f = 4$.

We now turn to hidden sector relics with $B \ll 1$. We integrate from $x_{\rm hid} = 2$ until both $\Phi$ and $f$ reach their asymptotic abundance. The resulting fractional energy densities are shown in the right panel of \Cref{fig:decoupled-freezeout}. As expected, the results are consistent with those from \Cref{sec:freezeout}. We numerically verify that both energy densities scale $\sim B^{0.28}$.

\textit{Coupled freeze-out.} We next consider the case in which $\lambda_\Phi m_{\Phi} = \lambda_f m_f \sim 3.7 \times 10^{16}\, \rm GeV$, but $\lambda_\Gamma = 1$. As before, we assume that both relics begin at their thermal abundance. In this case, we have a modified thermal freeze-out scenario -- the abundance of the $f$ particles at $x_{\rm vis} > 30$ allows the $\Phi$ particles to remain in equilibrium longer, via the inverse decay process $f \bar f \to \Phi$.

As in the previous case, we first integrate the visible sector model from $x_{\rm vis} = 2$ until $Y_f$ reaches its asymptotic value. Since the $\Phi$ relic is unstable and the decay term in the Boltzmann equation scales $\sim 1/T$, we have that $Y_\Phi\to 0$ as the Universe cools.

\begin{figure}
    \centering
    \includegraphics[width=\linewidth]{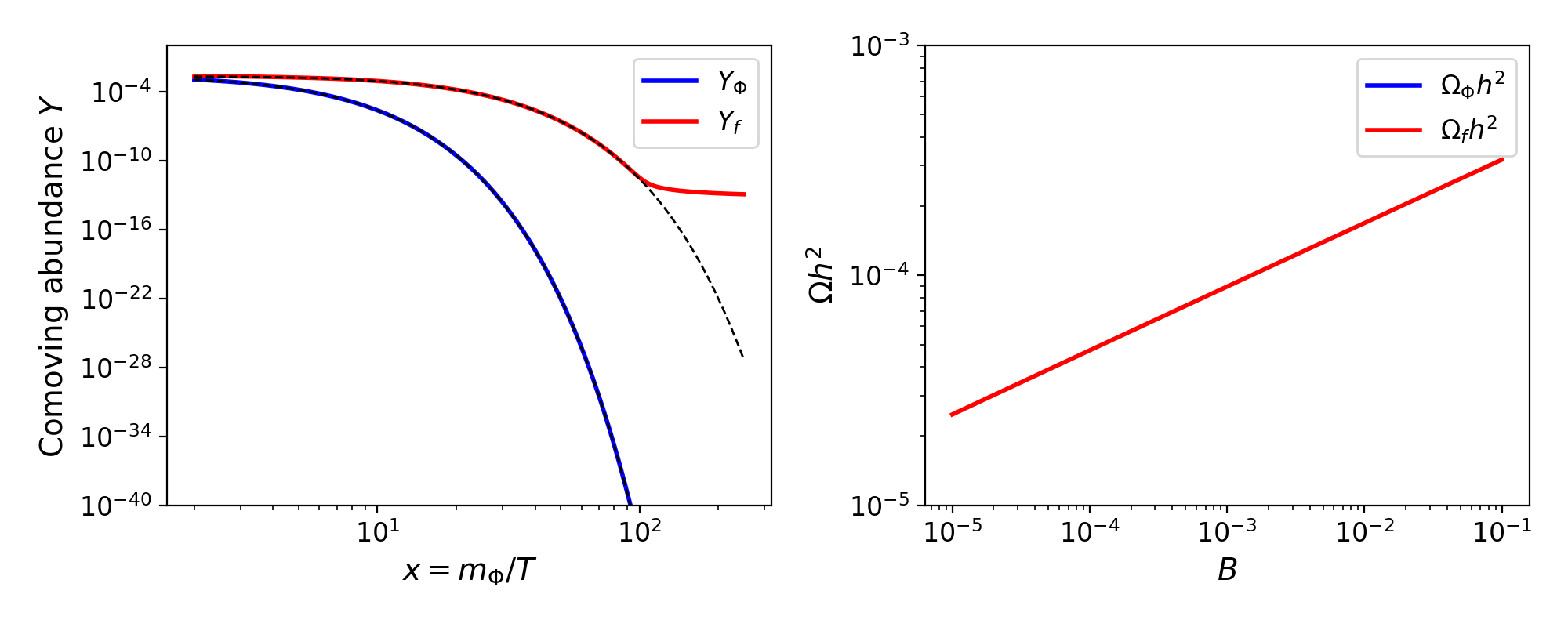}
    \caption{Left panel: Results of Boltzmann integration of coupled visible sector WIMPs, $\Phi$ (blue) and $f$ (red). Decoupling occurs at $x \sim 80$ for both $\Phi$ and $f$. Right panel: Dependence of the final energy density of $\Phi$ (blue) and $f$ (red) on the inflaton branching ratio $B$. The energy density of $f$ is proportional to $B^{0.28}$, in agreement with the case of single-component freeze-out; the energy density of $\Phi$ vanishes to numerical precision.}
    \label{fig:coupled-freezeout}
\end{figure}

The result of the integration is shown in the left panel of \Cref{fig:coupled-freezeout}. As expected, $\Phi$ remains in equilibrium much longer than in the previous example, since the process $f\bar f \leftrightarrow \Phi$ remains efficient. However, once the abundance of $f$ is frozen out at $x \sim 80$, $Y_\Phi$ can no longer maintain its equilibrium value. As the Universe cools, the remaining $\Phi$ particles decay into $f \bar f$ pairs. The abundance of $\Phi$ relics is sufficiently low ($\sim 20$ orders of magnitude less than $f$) so that $Y_f$ does not meaningfully increase.  

We now analyze the $B$-dependence of the corresponding hidden sector model. As before, we integrate from $x_{\rm hid}=2$ until the $f$ particle reaches its asymptotic abundance. The results are shown in the right panel of \Cref{fig:coupled-freezeout}. We numerically confirm that the asymptotic energy density of $f$ particles scales as $B^{0.28}$, consistent with the previous freeze-out scenarios. The asymptotic energy density of $\Phi$ particles vanishes due to rapid decay as the Universe cools. 

\textit{Coupled freeze-in.} Finally, we consider the case in which $\lambda_f m_f =  3.7\times 10^{16}\, \rm GeV$, $\lambda_\Phi=0$ and $\lambda_\Gamma = 10^{-8}$, and the $\Phi$ particle starts at a negligible abundance. Here, the $f$ particles will undergo ordinary freeze-out via annihilation, but annihilation of the $\Phi$ particles is impossible. The $\Phi$ particle will undergo freeze-in via the inverse decay process $f \bar f \to \Phi$.

As before, for the visible sector case, we integrate from $x_{\rm vis} = 2$ until $f$ reaches its asymptotic abundance. However, we assume $\Phi$ is metastable until the integration is halted at $x_{\rm vis}=250$. This ensures that the decay process $\Phi \to f \bar f$ remains inefficient at late times so that a meaningful $\Phi$ abundance is present at the end of the integration.

\begin{figure}
    \centering
    \includegraphics[width=\linewidth]{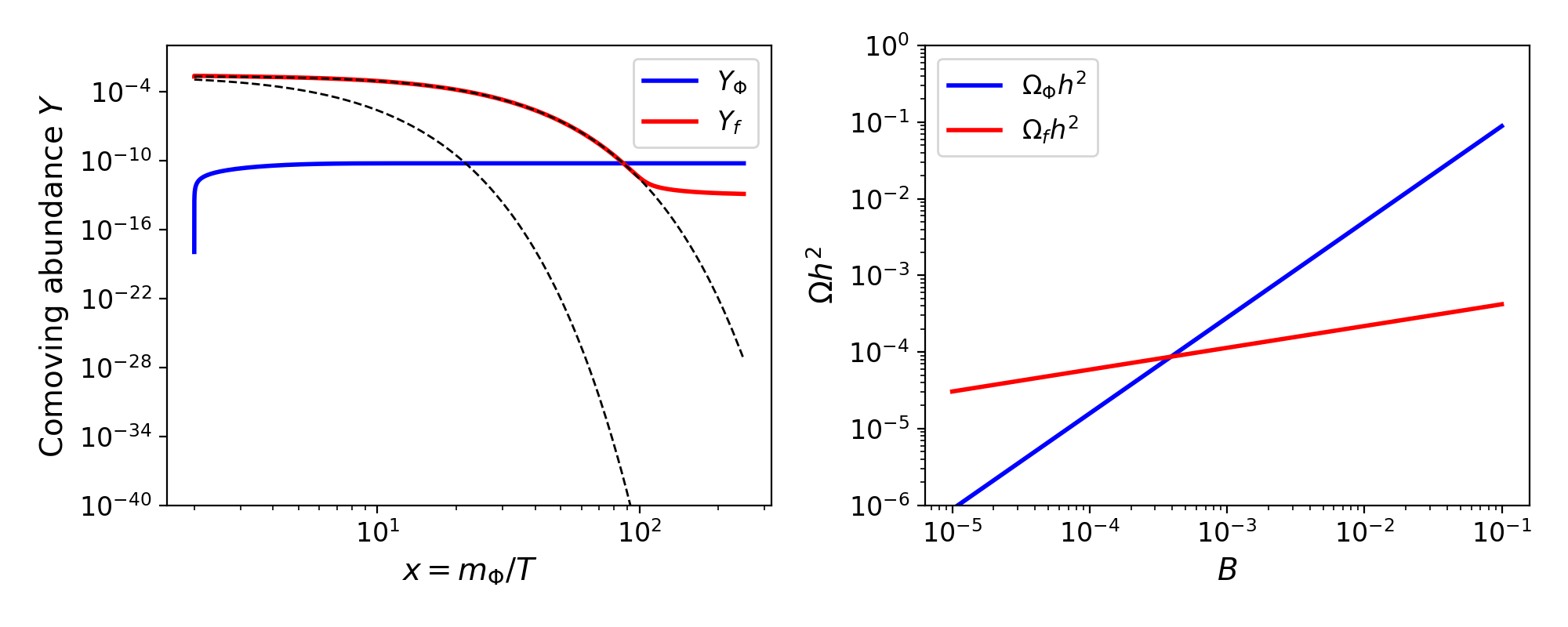}
    \caption{Left panel: Results of Boltzmann integration of coupled visible sector WIMPs, $\Phi$ (blue) and $f$ (red). Freeze-out occurs at $x \sim 80$ for the $f$ particles, and $\Phi$ is frozen in at high temperatures due to the inverse decay process $f \bar f \to \Phi$. Right panel: Dependence of the final energy density of $\Phi$ (blue) and $f$ (red) on the inflaton branching ratio $B$. The energy density of $f$ is proportional to $B^{0.28}$, in agreement with the case of single-component freeze-out; the energy density of $\Phi$ scales as $B^{1.25}$, in agreement with the single-component freeze-in results.}
    \label{fig:coupled-freezein}
\end{figure}

The results of the integration are shown in the left panel of \Cref{fig:coupled-freezein}. At early times, the inverse decay of the abundant $f \bar f$ pairs rapidly populates the $\Phi$ population to a sub-thermal abundance. This abundance is negligible compared to the equilibrium abundance of $f$ particles, so $Y_f$ does not meaningfully decrease. Afterwards, the $f$ relic undergoes standard freeze-out. We note that this integration intentionally truncates the low-temperature (high $x_{\rm vis}$) dynamics to preserve an abundance of $\Phi$, which would otherwise decay.

We now turn to the case of a hidden sector model and analyze the $B$-dependence of both relics. We integrate from $x_{\rm hid}=2$ to $x_{\rm hid}=250$ so that $\Phi$ decay remains inefficient for the entire integration. The resulting energy densities of the $\Phi$ and $f$ relics are shown in the right panel of \Cref{fig:coupled-freezein}. We numerically confirm that $\Omega_\Phi \sim B^{1.25}$ and $\Omega_f \sim B^{0.28}$, as expected from the single-component freeze-in results.

\subsubsection{Exchange DM}
We now consider the case of two stable scalars, $\phi_1$ and $\phi_2$ with mass $m_1$ and $m_2$, respectively. Each scalar may annihilate via the process $\phi_{1,2}\phi_{1,2}\to \rm radiation$, or exchange via the process $\phi_1 \phi_1 \leftrightarrow \phi_2 \phi_2$. This model is motivated by so-called \textit{exchange DM} in which the exchange of $\phi_1$ and $\phi_2$ can drive freeze-out \cite{maity2020exchange}.

The Boltzmann equations for this model are
\begin{align}
    \frac{dY_{1}}{dx} &= -\frac{\langle\sigma_{1}v\rangle s_{\rm vis}}{Hx}\left(Y_1^2 - Y_{1,\text{eq}}^2(xB^{-1/4})\right) -\frac{\langle\sigma_{12}v\rangle s_{\rm vis}}{Hx} \left(Y_1^2-\frac{Y^2_{1,\text{eq}}(xB^{-1/4})}{Y^2_{2,\text{eq}}(xB^{-1/4})}Y_{2}^2\right)\\
    \frac{dY_2}{dx} &= -\frac{\langle\sigma_{2}v\rangle s_{\rm vis}}{Hx}\left(Y_2^2 - Y_{2,\text{eq}}^2(xB^{-1/4})\right) + \frac{\langle\sigma_{12}v\rangle s_{\rm vis}}{Hx} \left(Y_1^2-\frac{Y^2_{1,\text{eq}}(xB^{-1/4})}{Y^2_{2,\text{eq}}(xB^{-1/4})}Y_{2}^2\right),
\end{align}
where $x=x_{\rm vis}=m_1/\Tv$, and $Y_1$ and $Y_{1, \rm eq}$ are the comoving abundance and equilibrium abundance of $\phi_1$, and likewise for $Y_2$, $Y_{2, \rm eq}$ and $\phi_2$. We stress that the equilibrium abundances are defined with respect to the visible sector entropy density. Additionally, $\langle\sigma_{1}v\rangle$ and $\langle\sigma_{2}v\rangle$ are the thermally-averaged annihilation cross sections for $\phi_1$ and $\phi_2$, and $\langle\sigma_{12}v\rangle$ is the thermally-averaged cross section for the process $\phi_1 \phi_1 \to \phi_2 \phi_2$. As before, we define
\begin{align}
    \lambda_1 &\equiv\frac{\langle\sigma_{1}v\rangle s_{\rm vis}}{H(\Tv=m_{\phi_1})} & \lambda_2 &\equiv\frac{\langle\sigma_{2}v\rangle s_{\rm vis}}{H(\Tv=m_{\phi_1})} & \lambda_{12} &\equiv\frac{\langle\sigma_{12}v\rangle s_{\rm vis}}{H(\Tv=m_{\phi_1})}
\end{align}
so that the Boltzmann equations become
\begin{align}
    \frac{dY_{1}}{dx} &= -\frac{\lambda_1}{x^2}\left(Y_1^2 - Y_{1,\text{eq}}^2(xB^{-1/4})\right) -\frac{\lambda_{12}}{x^2} \left(Y_1^2-\frac{Y^2_{1,\text{eq}}(xB^{-1/4})}{Y^2_{2,\text{eq}}(xB^{-1/4})}Y_{2}^2\right)\\
    \frac{dY_2}{dx} &= -\frac{\lambda_2}{x^2}\left(Y_2^2 -Y_{2,\text{eq}}^2(xB^{-1/4})\right) + \frac{\lambda_{12}}{x^2} \left(Y_1^2-\frac{Y^2_{1,\text{eq}}(xB^{-1/4})}{Y^2_{2,\text{eq}}(xB^{-1/4})}Y_{2}^2\right).
\end{align}

We now analyze the resulting abundances for visible and hidden sector instantiations of this model in various parameter regimes. For our analysis, we fix $m_1=100\, \rm GeV$ and $m_2=20\, \rm GeV$.

\textit{Decoupled freeze-out.} We first consider the case of decoupled freeze-out, in which $\lambda_1=3.6\times 10^{16}(\text{GeV}/m_1)$, $\lambda_2=3.6\times 10^{16}(\text{GeV}/m_2)$, but $\lambda_{12}=0$. Here, both $\phi_1$ and $\phi_2$ undergo freeze-out, and the process $\phi_1 \phi_1 \to \phi_2 \phi_2$ is suppressed. 

\begin{figure}
    \centering
    \includegraphics[width=\linewidth]{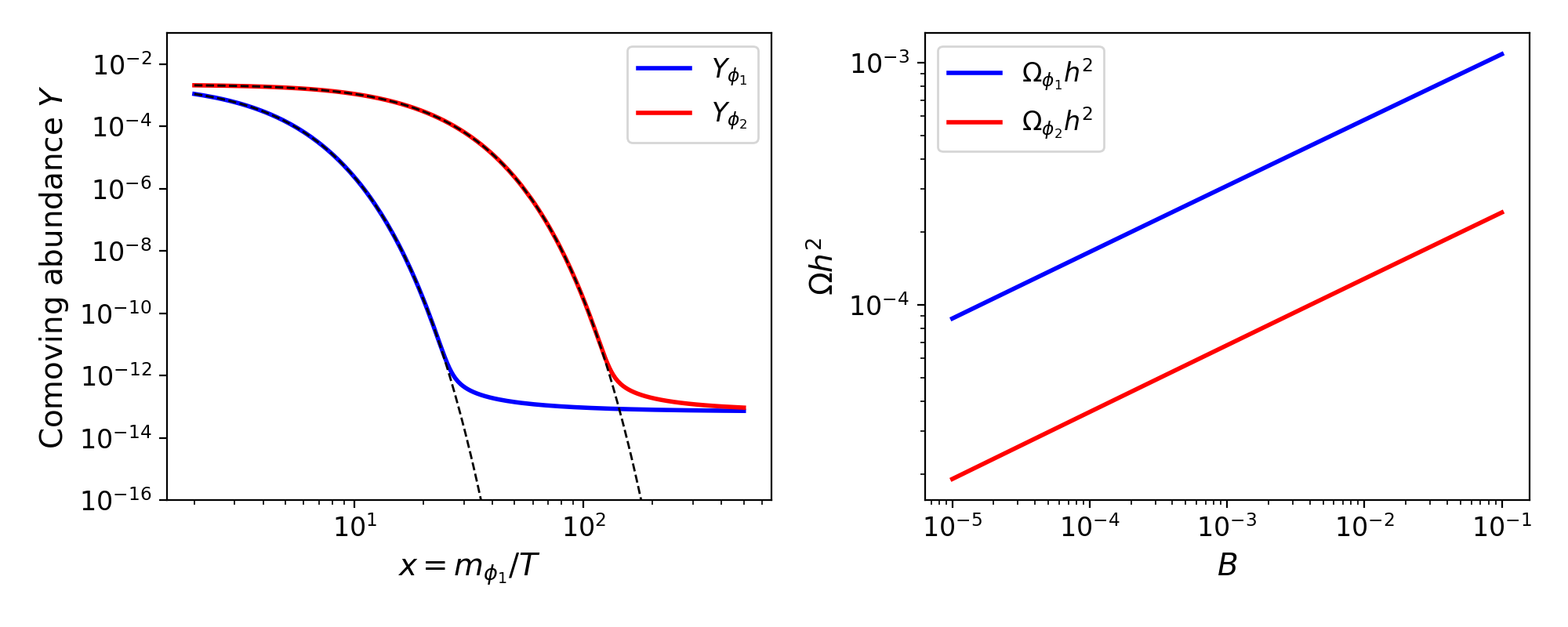}
    \caption{Left panel: Results of Boltzmann integration of decoupled visible sector WIMPs, $\phi_1$ (blue) and $\phi_2$ (red). Freeze-out occurs at $x \sim 20$ for $\phi_1$ and $x \sim 100$ for $\phi_2$. Right panel: Dependence of the final energy density of $\phi_1$ (blue) and $\phi_2$ (red) on the inflaton branching ratio $B$. Both energy densities are proportional to $B^{0.27}$, in agreement with the case of single-component freeze-out.}
    \label{fig:exchange-decoupled}
\end{figure}

The numerical results for the visible sector and hidden sector models are shown in \Cref{fig:exchange-decoupled}. As expected, the fractional energy densities of both $\phi_1$ and $\phi_2$ scale as $B^{0.27}$, consistent with our previous analytical and numerical results.

\textit{Assisted freeze-out.} Next, we consider the case where annihilation of $\phi_1$ is forbidden ($\lambda_1=0$), but $\phi_1$ may stay in equilibrium due to the process $\phi_2\phi_2 \to \phi_1 \phi_1$. We choose $\lambda_2=3.6\times 10^{16}(\text{GeV}/m_2)$ and $\lambda_{12}=10^{12}$.

\begin{figure}
    \centering
    \includegraphics[width=\linewidth]{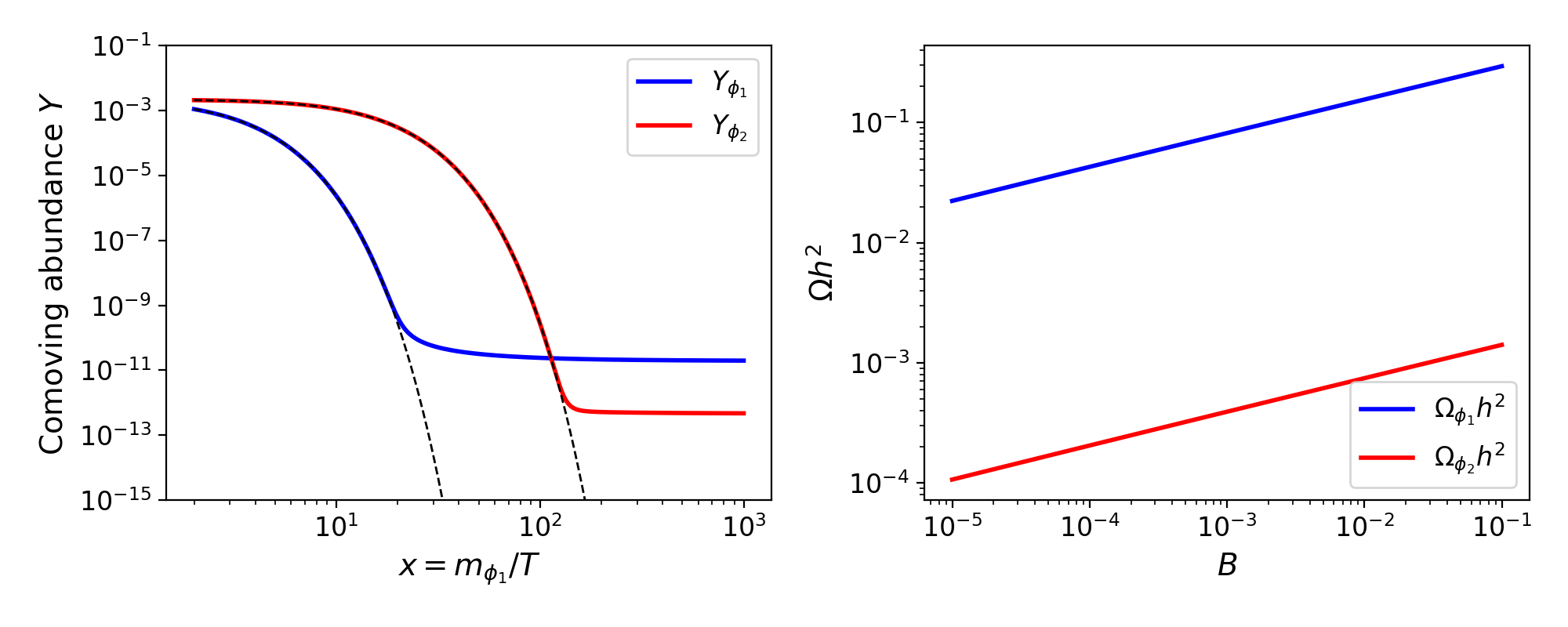}
    \caption{Left panel: Results of Boltzmann integration of coupled visible sector WIMPs, $\phi_1$ (blue) and $\phi_2$ (red). Freeze-out occurs at $x \sim 10$ for $\phi_1$ and $x \sim 100$ for $\phi_2$. Right panel: Dependence of the final energy density of $\phi_1$ (blue) and $\phi_2$ (red) on the inflaton branching ratio $B$. Both energy densities are proportional to $B^{0.28}$, in agreement with the case of single-component freeze-out.}
    \label{fig:exchange-freezeout}
\end{figure}

The numerical results for the visible and hidden sector models are depicted in \Cref{fig:exchange-freezeout}. The $\phi_2$ particle undergoes standard freeze-out, while $\phi_1$ undergoes assisted freeze-out until $x\sim 10$, when the production from $\phi_2 \phi_2 \to \phi_1 \phi_1$ becomes inefficient. This cutoff is due to exponential Boltzmann suppression $\sim e^{-2(m_1-m_2)/T}$. Again, the fractional energy densities of both $\phi_1$ and $\phi_2$ scale as $B^{0.28}$ for the hidden sector models.

\textit{Assisted freeze-in.} Finally, we consider the case in which $\phi_1$ begins at a subthermal abundance ($Y_1=0$) but is populated via the process $\phi_2 \phi_2 \to \phi_1 \phi_1$. As before, we take $\lambda_1 = 0$, $\lambda_2=3.6\times 10^{16}(\text{GeV}/m_2)$; however, in this case, we take a substantially smaller value of $\lambda_{12}=1$; this prevents $\phi_2 \phi_2 \to \phi_1 \phi_1$ from restoring $\phi_1$ to its equilibrium abundance.

\begin{figure}
    \centering
    \includegraphics[width=\linewidth]{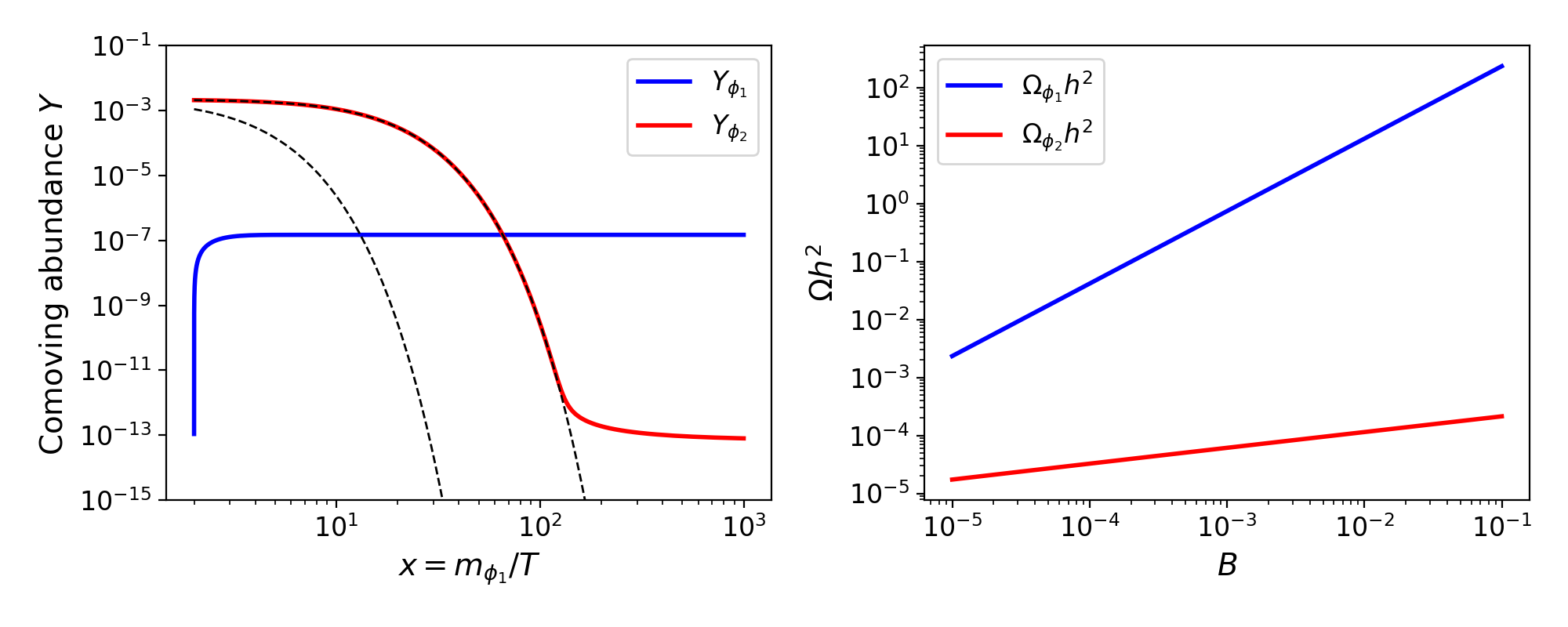}
    \caption{Left panel: Results of Boltzmann integration of coupled visible sector WIMPs, $\phi_1$ (blue) and $\phi_2$ (red). Freeze-out occurs at $x \sim 100$ for $\phi_2$, while $\phi_1$ is populated to a subthermal abundance via freeze-in. Right panel: Dependence of the final energy density of $\phi_1$ (blue) and $\phi_2$ (red) on the inflaton branching ratio $B$. As expected, the energy densities scale as $B^{1.25}$ and $B^{0.27}$ respectively.}
    \label{fig:exchange-freezein}
\end{figure}

The numerical results for the visible and hidden sector models are depicted in \Cref{fig:exchange-freezein}. As before, the $\phi_2$ particle undergoes standard freeze-out. However, in this case, $\phi_1$ is frozen in to a subthermal abundance by the process $\phi_2 \phi_2 \to \phi_1\phi_1$. The fractional energy densities of $\phi_1$ and $\phi_2$ scale as $B^{1.25}$ and $B^{0.27}$, respectively, consistent with our previous freeze-in and freeze-out results.

\section{Discussion}\label{sec:discussion}

The models considered in \Cref{sec:applications} are of the freeze-out and freeze-in type. In these models, the initial relic abundances are defined in terms of the hidden sector temperature and are independent of $B$; in the case of freeze-out it is set by thermality, while in the case of freeze-in, it is assumed to be zero. As discussed in \Cref{sec:general}, our formalism can also be applied to situations where the initial abundance of the relic has a non-trivial $B$-dependence. An important example of this is hidden sector monopoles produced by a second order phase transition. The initial hidden sector abundance immediately after the phase transition scales as $B^{-1/2}$ \cite{Graesser:2020hiv, Liveoak:2026muy}. Subsequently, the dilution is governed by a Boltzmann equation describing monopole-antimonopole pair annihilation. This equation can be solved analytically\footnote{The steps are very similar to the treatment in  \cite{Liveoak:2026muy}, which we do not detail here.}; the results are generally independent of the initial abundance and the final hidden sector abundance has an overall multiplicative factor of $B^{-1/2}$. Converting this to the abundance defined in terms of the visible sector one finds the $B^{1/4}$ dependence of \cite{Liveoak:2026muy}. We additionally note that hidden sectors which form monopoles but do not have light charged fermion states are generally expected to be dominated by the massive stable thermal dark gauge bosons $W'$ \cite{brummer2026no}. However, in this case, the $W'$ abundance is determined by thermal freeze-out and scales as $B^{1/4}$, thus making them dangerous from the perspective of overclosure.

We have focused on settings where the hidden sector is thermalized. It will be interesting to extend our analysis to more general situations. A key assumption of the effective Planck mass formulation is the fact that for a thermal hidden sector, the ratio of its energy density and the visible sector energy density remains constant at early times. This continues to hold for radiation-dominated non-thermal hidden sectors, but is not true if the hidden sector undergoes an era of matter domination. In such a situation, it is natural to incorporate a time-dependent effective Planck mass into our formalism. We note that such scenarios can have a major phenomenological challenge: any extended period of hidden sector matter domination will lead to a substantial increase in the hidden sector energy density relative to the visible sector energy density, which in turn will generally lead to a severe dark radiation problem.

As discussed in \Cref{sec:intro}, string compactifications generically predict the presence of multiple hidden sectors. It is interesting to consider the limit of a large number of sectors. Consider a setting with $N$ hidden sectors, with $\mathcal{B}$ the total branching ratio of the inflaton to the hidden sectors. The total DM energy density $\Omega_{\rm DM}$ is the sum of the individual contributions from each individual sector. If there are $M$ sectors with a freeze-out type thermal relic which each receive a branching ratio $\sim \mathcal{B}/N$, the DM energy density scales as $M (\mathcal{B}/N)^{1/4}$. Assuming a finite fraction of such sectors (i.e., $M/N$ finite in the limit of large number of sectors), the contribution scales as $N^{3/4}$, providing a large enhancement\footnote{On the other hand, an analogous analysis for sectors with freeze-in relics gives an $N^{-1/4}$ scaling, so the contribution diminishes with $N$.} that can significantly affect the phenomenology of such models.

Quasi-relics, which are long lived but unstable on cosmological timescales, can also have interesting implications. Their metastability may be attributed to weak coupling with a lighter particle that shares a conserved quantum number. These couplings may connect two or more hidden sectors. Our results imply that freeze-out quasi-relics may be produced in abundance, even if the inflaton has a low branching ratio to the sector(s) they reside in. They will eventually decay to the lightest particle with the conserved quantum number, leading to enhancements in its number density.

  Supersymmetric theories naturally provide WIMP dark matter with its mass governed by
the supersymmetry breaking scale and mediation mechanism. Given a mediation mechanism, our results can be translated to bounds on soft masses/parameters. This should provide interesting constraints on supersymmetric theories with hidden sectors.

\section{Conclusion}\label{sec:conclusion}

In this paper, we have developed a general framework to study the DM relic density from thermalized hidden sectors. The contribution of a hidden sector relic to the DM abundance is determined by microscopic dynamics of the relic and the fraction of the total energy of the Universe deposited in the sector, which is determined by the inflaton branching ratio to the sector $B$. The dependence of the DM energy density on $B$ is crucial for determining the impact of a sector on cosmology -- sectors with a weak dependence can contribute significantly to DM (or lead to overclosure) even if they carry a small fraction of the total energy. Our analysis of the Boltzmann equations has shown that this dependence can be captured by defining an effective Planck mass. We have applied our formalism to a large class of models, including freeze-out/freeze-in WIMPs, multi-component dark matter, and dark monopoles. The dependence on the branching fraction is weak $(\sim B^{1/4})$ for single- and multi-component freeze-out models and dark monopoles produced via second order phase transitions. These relics can contribute significantly to the DM energy density even if the inflaton deposits a small fraction of its energy to the associated hidden sector. On the other hand, usual freeze-in dark matter exhibits a much stronger dependence $(\sim B^{5/4})$. The case of production of weakly interacting particles produced by UV dominated non-renormalizable interactions is an exception, the dependence is even stronger ($\sim B^{3/2}$).

Hidden sectors arise naturally in BSM constructions and models of particle physics from string theory. This paper provides the necessary tools to determine the contribution of hidden sector(s) to DM once the fractional energy density of the sectors is determined from inflation (or any alternate theory of early Universe dynamics). The weak dependence on the inflaton branching ratio for a large class of models implies that relics from such sectors can make significant contributions to the DM energy density. Our results are important for constraining the space of phenomenologically viable hidden sector models on the basis of DM overabundance.

\begin{acknowledgments}
We thank T. Xie for correspondence that led to correcting earlier versions of \Cref{fig:wimp-freezeout} and \Cref{fig:decoupled-freezeout}. DL is supported by the National Science Foundation Graduate Research Fellowship Program. AM would like to thank the Leinweber Institute for Theoretical Physics for supporting his sabbatical visit at the University of Michigan, Ann Arbor. JW acknowledges support from the Leinweber Foundation.
\end{acknowledgments}


\bibliography{apssamp}

\end{document}